\documentclass[a4paper,fleqn,usenatbib]{mnras}

\usepackage{newtxtext,newtxmath}

\usepackage[T1]{fontenc}
\usepackage{ae,aecompl}
\usepackage[utf8]{inputenc}

\usepackage{graphicx}	
\usepackage{amsmath}	
\usepackage{amssymb}	

\usepackage{float}
\usepackage{color}
\usepackage{booktabs}
\usepackage{multirow}
\usepackage{siunitx}
\usepackage{multicol}
\PassOptionsToPackage{pdfpagelabels=true}{hyperref}
\usepackage{lastpage}

\title[Inner view of NGC\,1052]{The inner view of NGC\,1052 using multiple X-ray observations}

\author[N. Osorio-Clavijo et al.]{
N. Osorio-Clavijo,$^{1}$ \thanks{email: n.osorio@irya.unam.mx}
O. González-Martín,$^{1}$
I. E. Papadakis,$^{2,3}$
J. Masegosa,$^{4}$
\newauthor
L. Hernández-García$^{5}$
\\
$^{1}$Instituto de Radioastronomía and Astrofísica (IRyA-UNAM), 3-72 (Xangari), 8701, Morelia, Mexico \\
$^{2}$Department of Physics \& Institute of Theoretical \& Computational Physics, University of Crete, PO Box 2208, 710 03 Heraklion, Crete, Greece \\
$^{3}$IESL, Foundation for Research and Technology-Hellas, 71110 Heraklion, Crete, Greece \\
$^{4}$ IAA – Instituto de Astrofísica de Andalucía (CSIC), Apdo. 3004, E-18080 Granada, Spain\\
$^{5}$ Instituto de Física y Astronomía, Facultad de Ciencias, Universidad de Valparaíso, Gran Bretaña 1111, Playa Ancha, Valparaíso, Chile
}
\date{Accepted 2019 September 30. Received 2019 September 26; in original form 2019 April 3.}

\pubyear{2019}
\begin{document}
\label{firstpage}
\pagerange{\pageref{firstpage}--\pageref{lastpage}}
\maketitle


\begin{abstract}
In this paper, we make a multi-epoch analysis of NGC\,1052, one of the prototypical LLAGN, using \textit{XMM}-Newton, \textit{Suzaku} and \textit{NuSTAR} observations, taken from 2001 to 2017. This is the first time that results from \textit{NuSTAR} observations are reported for NGC\,1052. {On the technical aspects, we found a wavelength-dependent calibration issue between simultaneous \emph{XMM}-Newton and \emph{NuSTAR} spectra. It is described by a change on the photon index of $\rm{ \Gamma_{NuSTAR}- \Gamma_{XMM-Newton}=0.17\pm0.04}$.} We use ancillary \emph{Chandra} data to decontaminate the nuclear spectrum from circumnuclear contributors. \ We find that two baseline models can fit the broad (0.5-50 keV) X-ray spectrum of the source. One consists of a power-law like continuum which is absorbed by a uniform absorber, reflection from neutral material, and a separate power-law component in the soft band. The second model presents a clumpy absorber. The reflection component is still present, but not the soft band power-law. Instead, absorption by a warm absorber is necessary to fit the spectra. This is the first time that a reflection component is established in this object, thanks to high energy data from \emph{NuSTAR}. This component is constant in flux and shape, supporting the idea that is produced away from the central source (probably the torus). We find flux, spectral slope and absorption variations on timescales of months to years. We also find that a patchy-absober can explain the behaviour of this source better as it is $\sim$ 200 times more likely than the uniform absober while it yields to smaller intrinsic variations.
\end{abstract}




\section{Introduction} \label{intro}

Active Galactic Nuclei (AGN) are one of the most energetic phenomena in the Universe, with luminosities that can go up to the $\rm{10^{48} \,erg \ s^{-1}}$. This release of energy is thought to be due to the feeding of the super massive black hole (SMBH) residing at the center of every galaxy \citep[][]{Peterson-97} by an accretion disc. The AGN unified model \citep[UM,][]{Antonucci-93, Urry-95} consists of a SMBH surrounded by the accretion disc, an obscuring dusty torus, both narrow- and broad-line regions (so called NLR and BLR, respectively) and a radio jet. Different AGN are seen depending on the point of view to the observer, hence the AGN classification may be purely due to a geometrical effect. However, although a substantial amount of AGN are explained via the UM, there are some types of objects not easily explained (e.g., the Low-Luminosity AGN) and the intrinsic power of the AGN is likely to play an important role. Therefore, several modifications of this UM have been proposed, some of which include: i) to modify the morphology of the obscuration, ii) to review the accretion mechanism and/or the nature of the central source which might explain the absence of the NLR or the BLR and iii) to take into account the co-evolution of the SMBH and the host galaxy \citep[see][for a review]{Netzer-15}. 

Low-Luminosity AGN (LLAGN), which dominate the population of nearby AGN \citep{Ho-97}, constitute a type of object which cannot be explained by the standard UM due to their sub-Eddington luminosities ($L_{2-10 \ \rm{keV}} \approx 10^{39}-10^{42} \ \mathrm{erg \ s^{-1}}$ \citealt{Brenneman-09}). 
\cite{Ho-08} suggested that they are not a simply scale-down version of other AGN, and that their emission processes must be different that the conventional mechanisms invoked for more luminous AGN. Some of the possibilities to explain their low luminosities include inefficient radiation of the accretion flow \citep[RIAF,][]{Narayan-05}, heavy absorption of the nuclear emission \citep{Gonzalez-09}, a combination of both \citep{Brenneman-09}, or the fact that the SMBH is being fed by stellar winds \citep{Heckman-14}.

It was first stated by \citet{Maoz-05}, using UV data, that LLAGN are variable sources; these variations are in time-scales ranging from months to years \citep{Gonzalez-12} and its variability pattern has been extensively studied at X-rays by \citet{Hernandez-13, Hernandez-14}. However, the most energetic part of the X-ray spectrum has barely been taken into account due to the lack of observations. The spectrum above 10 keV is mandatory to disentangle the intrinsic continuum, the intrinsic obscuration, and the reflection component. Although there are some dedicated works to study the high energy spectrum of LLAGNs \citep[e.g.][]{Younes-19}, very few have taken into account data above 10 keV to study long-term variations \citep[see e.g.][]{Gonzalez-11a}.

Our purpose is to perform an analysis of the long-term variations in the LLAGN NGC\,1052, classified as a LLAGN \citep{Ho-97}, with $\rm{L_{bol}} = 6.92 \times 10^{43} \ \mathrm{erg \ s{-1}}$ \citep{Brenneman-09}. This is one of the most studied archetypes of LLAGN, with a hidden BLR found by \cite{Barth-99}. \citet{Guainazzi-99} were one of the firsts to study NGC\,1052 at X-rays and catalogued it as an exceptional Low Ionization Narrow Emission Region (LINER).  

NGC\,1052 is a well known variable source at X-rays, however no consensus has been achieved about what is causing these variations. Two possible scenarios have been proposed: i) an obscuring structure along the line of sight \citep{Hernandez-13} and ii) a variable accretion disc \citep{Connolly-16}. In this work, we find for the first time, variations in both obscuration throughout the hydrogen column density, $\rm{N_H}$, and the shape of the energy distribution of photons, i.e. the photon index, $\rm{\Gamma}$. To do this, we use data from \textit{XMM}-Newton, \textit{Suzaku} and \textit{NuSTAR}, from 2001 to 2017 and fit them simultaneously by using both soft and hard energy bands with a model that accounts for non-negligible reflection component at energies above 10 $\mathrm{keV}$. {In fact, this is the first time that such a component has been shown to be present in the X-ray spectrum of this source.} We use ancillary \textit{Chandra} observations to study extended contributors that might be contaminating the nuclear spectrum.

This paper is organized as follows: section \ref{Data} introduces
the data and the data reduction; in section \ref{data-analysis} we motivate the baseline model used to fit the data, and in section \ref{discussion} we present our results. Finally, in section \ref{summary} we summarize our work and we propose a scenario for this variable source. 

\begin{table}
\label{tab:observaciones}
\centering
	\begin{tabular}{lcccc} 	\hline
		Satellite & ObsID  & Date  & Expt. & Apert.  \\
		 &  &   &  ($ks$) & (arcsec) \\  \hline
	 	\textit{XMM} & $0903630101$ (X1) & $2001$-$08$-$15$  & $16$ & 25  \\
	 	\textit{Chandra*} & $5910$ & $2005$-$09$-$18$ & $60$ & 3-25 \\
	 	\textit{XMM}  & $0306230101$ (X2) & $2006$-$01$-$12$ & $55$ & 25 \\
	 	\textit{Suzaku} & $702058010$ (S) & $2007$-$07$-$16$ & $101$ & 144 \\
	 	\textit{XMM}  & $0553300301$ (X3) & $2009$-$01$-$14$ & $52$ & 25  \\
	 	\textit{XMM}  & $0553300401$ (X4) & $2009$-$08$-$12$ & $59$ & 25  \\
	 	\textit{NuSTAR} & $60061027002$ (N1) & $2013$-$02$-$14$ & $15$ & 64  \\ 
	 	\textit{NuSTAR} & $60201056002$ (N2) & $2017$-$01$-$17$ & $60$ & 64  \\
	    {\textit{XMM}} & $0790980101$ (X5) & $2017$-$01$-$17$ & $70$ & 25 	\\ \hline
	\end{tabular}
\caption{Data used in this work. Fourth and fifth columns list the net exposure time and the aperture size we used to extract the data, respectively. Note that in the case of \textit{Chandra} data we used a ring to decontaminate nuclear from circumnuclear contributors (see Section \ref{data-red} for details).}

\end{table}

\begin{figure*}
    \centering
    \includegraphics[width=2.0\columnwidth]{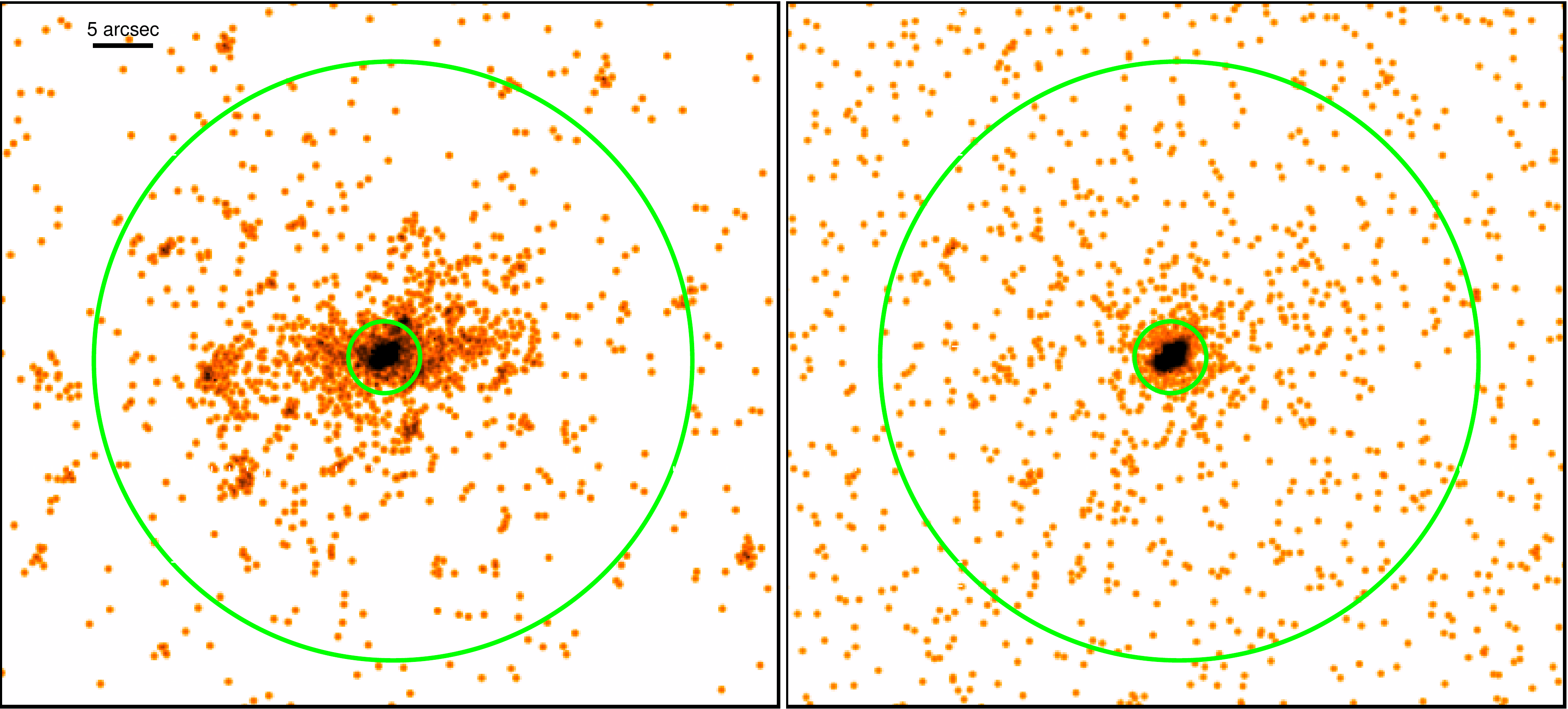}
    \caption{0.5-2.0 keV and 2.0-10.0 keV  \textit{Chandra} images (left and right panels, respectively). Green circles indicate the ring we used to extract circumnuclear spectrum.}
    \label{Chandra}
\end{figure*}

\section{Data}
\label{Data}

NGC\,1052 (RA (J2000)= 02:41:04.8 and Dec (J2000) = -08:15:21) has been observed with most of the current and previous X-ray missions (i.e., \emph{BeppoSAX}, \textit{XMM}-Newton, \emph{Chandra}, \emph{Suzaku}, \textit{NuSTAR}, \emph{RXTE}, and \emph{Swift}). In this work, we use data from \textit{XMM}-Newton, \emph{Chandra}, \emph{Suzaku} and \textit{NuSTAR}. Tab.\,\ref{tab:observaciones} contains relevant information on these observations: the observation ID, the observing starting date, the net exposure time, and the aperture size used to extract each spectrum.

We use data from all the public observations from these missions, except one \emph{XMM}-Newton observation, due to its short exposure time. All together we collected seven observations distributed in 16 years, from 2001 up to 2017. The shortest time-scale we can probe is $\sim$7 months. {Note that the latest \emph{NuSTAR} observation (N2 in Table\,\ref{tab:observaciones}) is simultaneous with the latest \emph{XMM}-Newton observation (X5 in Table\,\ref{tab:observaciones}).}

We downloaded data from the HEASARC archive\footnote{\url{https://heasarc.gsfc.nasa.gov/}}. In all cases, we identified the source position using the coordinates from NED\footnote{\url{https://ned.ipac.caltech.edu}}. The aperture radius for the different satellites was chosen as a compromise between achieving the maximum S/N and ensuring that we encompass $>80 \%$ of the point spread function (PSF) of each instrument.

\subsection{Data reduction} 
\label{data-red}

$\rm{\bullet}$ \underline{\emph{Chandra}} (0.5-8 keV): \emph{Chandra} \citep{Weisskopf-02} data were  analysed using the \emph{CXC Chandra Interactive Analysis of Observations} (CIAO\footnote{\url{http://asc.harvard.edu/ciao}}) software version 3.1, following standard procedures \citep[e.g.][]{Gonzalez-09}. The exposure time was processed to exclude background flares, using the task {\sc lc\_clean.sl} \footnote{\url{http://cxc.harvard.edu/ciao/download/scripts/}} in source-free sky regions of the same observation.

Due to its superior spatial resolution, we used the \textit{Chandra} data to study the circumnuclear emission of the source. We did not use \emph{Chandra} nuclear spectrum due to the low S/N ratio (see Section \ref{circumnuclear}). We extracted an annular region of inner radius 3 arcsec (to avoid the nuclear emission) and outer radius of 25 arcsec (this is equal to the \textit{XMM}-Newton spectrum extraction radius, see below).

Figure \ref{Chandra} shows the soft and hard band \textit{Chandra} images with the 25 arcsec circular regions marked as the large green circles. These images have been automatically smoothed with a Gaussian filter to enhance extended emission. Extended emission, beyond the central source, is easily identified in the soft band. Most of it is associated with diffuse emission, although some point-like sources can also be seen. The \textit{Chandra} images clearly show that diffuse emission cannot be detected at distances larger than 25 arcsec. In the next section we analyse the spectrum of the extended emission obtained with the \textit{Chandra} data, in order to account for it in the spectra extracted from the other instruments (which have worse spatial resolution than \textit{Chandra}).

\noindent $\rm{\bullet}$ \underline{\emph{XMM}-Newton} (0.5-10 keV): \emph{XMM}-Newton data used in this analysis are from the EPIC pn camera \citep{Struder-01}. The data were reduced with SAS v15.0.0, using the most up-dated available calibration files and  we followed standard procedures \citep[e.g.][]{Gonzalez-12}. 

 We used circular regions with 25 arcsec radii (500 pixels) to extract spectra of the target. The background events were selected from a source-free circular region on the same CCD as the source. We selected only single and double pixel events (i.e. PATTERN==0-4). Bad pixels and events too close to the edges of the CCD chip were rejected (using the standard FLAG==0 inclusion criterion). Background flares (periods of enhanced count rate) were removed by extracting the light curve of a source-free region and excluding time intervals with the background above three times the average count-rate of the light curve.

\noindent $\rm{\bullet}$ \underline{\emph{Suzaku}} (0.7 - 30 keV): For \emph{Suzaku} \citep{Mitsuda-07} data, we used the X-ray Imaging Spectrometer (XIS) that has an energy coverage from 0.7 - 7.6 keV and the Hard X-ray Detector (HXD) that has an energy coverage from 10 - 30 keV. The data are observed at the HXD nominal point to maximise the S/N. For the data reduction and analysis we followed the latest \textit{Suzaku} data reduction guide\footnote{\url{http://heasarc.gsfc.nasa.gov/docs/suzaku/analysis/ abc/}}. The extraction of the source was made by using a circular region of 2.4 arcmin radius. We reprocessed all the data files using standard screening within XSELECT ($\rm{SAA==0}$ and $\rm{ELV > 5}$). We reprocessed the spaced-row charge injection (CTI) data of the XIS instrument using {\sc xispi} task in order to use the latest calibration files at the time of writing. We also excluded data with Earth day-time elevation angles below 20{\textdegree} using XSELECT ($\rm{DYE\_ELV > 20}$). The XIS data were selected in 3$\rm{\times}$3 and 5$\rm{\times}$5 edit-modes using grades 0, 2, 3, 4, 6. Hot and flickering pixels were removed using the {\sc sisclean} script.

\noindent $\rm{\bullet}$ \underline{\emph{NuSTAR}} (3-50 keV): \emph{NuSTAR} \citep{Harrison-13} data were reduced using the data analysis software \emph{NuSTARDAS} v.1.4.4 distributed by the High Energy Astrophysics Archive Research Center (HEASARC). The calibrated, cleaned and screened event files for both FPMA and FMPB focal plane modules were generated using the {\sc nupipeline} task (CALDB 20160502). A circular region of $\sim1$ arcmin radius was taken to extract the source and background spectrum on the same detector and to compute the response files (RMF and ARF files) using the {\sc nuproducts} package available in \emph{NuSTARDAS}. Finally, we used the {\sc grppha} task within the FTOOLS to group the spectra with at least 60 counts per bin. 

We searched and we found that pile-up is not significant in any of the observations. Finally, \emph{XMM}-Newton images do not show any other source within the \textit{Suzaku} and \textit{NuSTAR} aperture radius, so any additional contributions to these spectra from nearby, individual sources must be negligible.

\section{Spectral fitting results} 
\label{data-analysis}

Spectral analysis was performed with {\sc xspec}\footnote{ \url{http://cxc.heasarc.gsfc.nasa.gov/docs/xanadu/xspec/}} (version 11.3.2), using $z = 0.00508$ (from NED). 
Errors in the best-fit parameters and in the flux measurements correspond to 1$-\sigma$ errors. In the following, we accept that a model gives a good fit to the data if the \textit{p}-value is larger than 0.01 and that more complex models provide a better fit to the data if the F-statistic probability is less than $10^{-3}$. Whenever two model versions have the same degrees of freedom, $\rm{dof}$, we use the `evidence ratio', $\rm{\epsilon_2}$,  in Eq. 8 of \cite{Emmanoulopoulos-16} to compare the goodness-of-fit. If $\rm{\epsilon_2} < 0.01$, the model with a smaller $\chi^2$ will be preferred.

\begin{figure}
	\includegraphics[width=1.0\columnwidth]{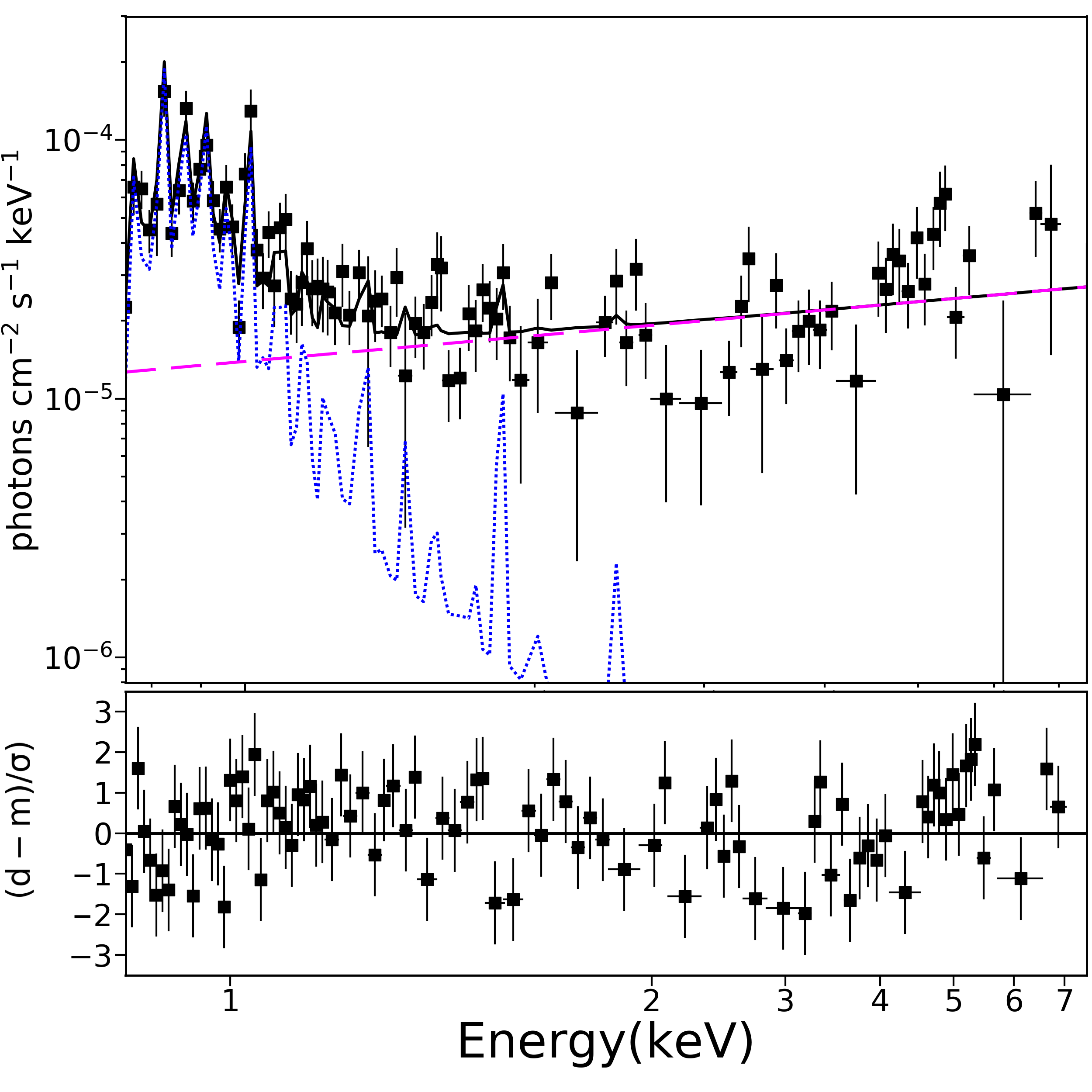}
    \caption {Spectral fit (top) and residuals (bottom) to the circumnuclear \emph{Chandra} data. Black solid, blue dotted and pink long-dashed lines in the top panel show the best-fit, thermal, and power-law components, respectively (see Section \ref{circumnuclear})}.
    \label{circum}
\end{figure}

\subsection{Circumnuclear emission spectrum} 
\label{circumnuclear}

Prior the simultaneous fit to all the spectra of NGC\,1052, we investigated how extended emission contributes to the total emission at the \textit{XMM}-Newton, \textit{Suzaku}, and \textit{NuSTAR} apertures. We fit the \emph{Chandra} circumnuclear (3-25 arcsec) spectrum to a combination of thermal (i.e., {\tt mekal}) and power-law components, applying Galactic absorption as well. We kept the {\tt mekal} parameters (i.e., hydrogen column density, abundance, and switch at their default values, 1, 1, and 1), and we let the temperature and normalization free to vary. Our best-fit values are $\rm{kT =0.62\pm0.04\ keV}$ and $\rm{\Gamma = 0.7\pm 0.2}$, and the model fits the data well ($\rm{\chi^2/dof} = 143.17/114$), where $\rm{dof}$ stands for degrees of freedom; $p$-value = 0.03. The circumnuclear spectrum and the best-fit components are shown in Fig.\,\ref{circum}. 

The {\tt mekal} component accounts for thermal emission from the host galaxy. The best-fit temperature is in agreement with other works \citep[e.g.][]{Gonzalez-09, Brenneman-09, Hernandez-13}. The origin of the power-law component is less clear; it might be associated with a jet component, reported in the literature \citep[][]{Kadler04b}, although it may also include emission from point-like sources (probably X-ray binaries). In the following, we use this spectral fit (with all the parameters fixed) when we fit the \textit{XMM}-Newton, \textit{Suzaku}, and \textit{NuSTAR} data. 

\subsection{\textit{XMM}-Newton and \textit{NuSTAR} cross-calibration}\label{sec:mismatch}

\begin{figure}
\includegraphics[width=1.0\columnwidth]{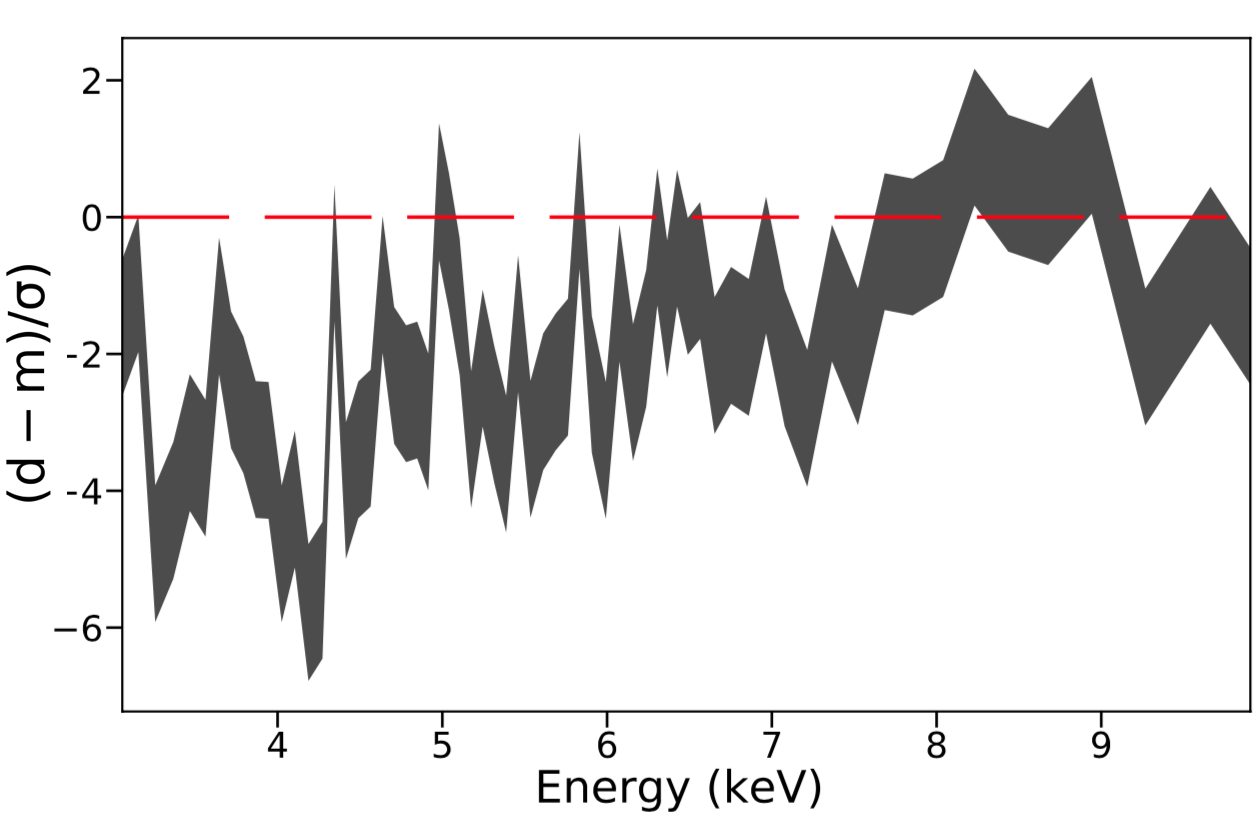}
\caption{Residuals between \emph{XMM}-Newton X5 observation and \emph{NuSTAR} N2 best-fit model.}
\label{xmm}
\end{figure}

{We take the advantage of the simultaneous observations \textit{NuSTAR} N2 and \textit{XMM}-Newton X5 (see table \ref{tab:observaciones}) to investigate any cross-calibration issue between these two instruments, as it has been suggested in the past  \citep[e.g.][]{Marchesi-18}. Indeed, if not properly taken into account, this issue might lead to a wrong estimate of the variations in the photon index, and in general, to a wrong estimate on the intrinsic parameters. 
We firstly fitted the \textit{NuSTAR} N2 with the combination of an absorbed power-law component plus a reflection component ({\tt pexmon}, see section\,\ref{refl}), with the photon indices of both components linked. We obtained $\chi^2/\rm{dof} = 660.7/625$,  $\rm{\Gamma_{HB}} = 1.80\pm 0.03$ and $\rm{N_H = 12.78 \pm 0.60\times 10^{22} \ cm^{-2}}$. \\

Figure \ref{xmm} shows the model residuals when the best-fit is applied to the \textit{XMM}-Newton X5  spectrum. This plot shows a clear wavelength-dependent calibration issue, suggesting that the simultaneous \emph{NuSTAR} and \emph{XMM}-Newton spectra cannot be explained with the same spectral slope. For this reason, we refitted the \emph{XMM}-Newton X5 spectrum, keeping the {\tt pexmon} component frozen to the parameters obtained with \emph{NuSTAR} and first letting the normalization of the power-law free, obtaining ($\rm{\chi^2/dof} = 412.44/421)$. We also tried the scenario in which the normalization is linked and $\rm{\Gamma_{HB}}$ is free, obtaining $451.75/421$. This is not better than the previous case ($\epsilon_2 \sim 3\times 10 ^{-9}$). Finally, we let both parameters free to vary. We obtained a $\rm{\Delta \chi^2 = 26.1}$ for one extra dof, and $\rm{\Gamma_{HB}} = 1.64\pm0.03$. The F-test probability ($10^{-7}$) implies that this scenario is statistically preferred. This result suggests that that the difference of $\rm{\Gamma_{NuSTAR} - \Gamma_{XMM-Newton}= 0.17\pm 0.04}$ between \textit{XMM}-Newton and \textit{NuSTAR} spectra, is significant.}

{A deeper investigation of this isse is out of the 
scope of this paper. In the subsequent analysis, we will apply this correction factor to the values calculated for all \textit{NuSTAR} (i.e., we report the \textit{NuSTAR} values corrected by 0.17). Note that, although the corrections have been made on the \textit{NuSTAR} data, the reader should be warned of the fact that all photon indices might be stepper by a factor of 0.17, because we cannot state which instrument is over/underestimating the parameter.}

\subsection{Spectral models for the hard band spectra}
\label{sec:spectralmodels}

{For the hard band (i.e., energies above 3 keV) we assumed two models: (i) a uniformly obscured intrinsic power-law with a high energy cutoff and (ii) a partially covered power-law with a high energy cutoff. Both are combined with a reflection component (see Sec. \ref{sec:reflection}). The baseline model equation, in {\sc xspec} terminology, for the uniform absorber is:}
\begin{equation}
\label{uniform-absorber}
M_1(E) = {\tt phabs_{Gal}}*({\tt phabs_{intr}}*{\tt zcutoffpl_{HB}} + {\tt pexmon}  )
\end{equation}

\noindent {where ${\tt phabs_{intr}}*{\tt zcutoffpl_{HB}}$ is the uniformly absorbed intrinsic continuum. In the case of partial-covering continuum we use ${\tt pcfabs_{intr}}*{\tt zcutoffpl_{HB}}$. Therefore, the baseline model is:}
\begin{equation}
\label{partial-covering}
M_2(E) = {\tt phabs_{Gal}}*({\tt pcfabs_{intr}}*{\tt zcutoffpl_{HB}}+{\tt pexmon}  )
\end{equation}
{In both scenarios, the component ${\tt pexmon}$ \citep{Nandra-07} accounts for reflection from neutral material. The free parameters of the combined model are, in both cases:} the intrinsic absorption, $\rm{N_{H_{intr}}}$, the hard band power-law photon index, $\Gamma_{\rm{HB}}$, its normalization, the energy cutoff, $\rm{E_{cutoff}}$ (for both ${\tt zcutoffpl}$ and ${\tt pexmon}$ components) the photon index of the reflection component, $\rm{\Gamma_{pex}}$, which might or not be related to that of the continuum, and its normalization. {In the case of the partial-covering scenario, the covering fraction $\rm{C_f}$ is also a free parameter.} All the other parameters of the ${\tt pexmon}$ component were kept fixed as follows: abundance (in solar units, 1.0), relative reflection (-1.0), iron abundance (1.0), and the inclination ($85$\textdegree). We did some tests varying these parameters but the data seem to be insensitive to them. Note that in all fits, we assume that the energy cutoff of the power-law and of the reflection component are linked, i.e., $\rm{E_{cutoff,pex}}$ =$\rm{E_{cutoff,zcutoff}}$. We tried to leave this parameter free to vary but it was not constrained. Therefore, we fixed it to 300 keV throughout all the analysis.

\subsection{Reflection component}
\label{refl}
We investigate whether this component is present in the spectrum of the source by using the \textit{Suzaku} and \textit{NuSTAR} observations only, because they let us access to energies above 10 keV, mandatory to study the presence of the reflection component. 
{Since it is the first time that a reflection component is reported with high significance for NGC\, 1052, we give particular care to verify its presence in this section.}
For this purpose, we  simultaneously fit the hard band spectra {using the baseline models above with and without the reflection component. Without the reflection component we obtain with $\rm{M_1}$ a $\chi^2/\rm{dof} = 1154.2/1049$ with $\rm{N_H = 11.0\pm 0.4\times 10^{22} \ cm^{-2}}$ and $\rm{\Gamma_{HB}}=$ $1.41\pm 0.04$, where all photon indices are tied together\footnote{In reality, we add to the photon indices of \textit{NuSTAR} t$\Delta \Gamma$ difference found above, assuming a mismatch between \textit{NuSTAR} and \textit{Suzaku}, similar to the one between \textit{NuSTAR} and \textit{XMM}-Newton.}
For $\rm{M_2}$ we obtain $\rm{\chi^2/\rm{dof} = 1099.2/1048}$, $\rm{N_H = 24.3\pm 1.8 \times 10^{22} \ cm^{-2}} $ and $\rm{\Gamma_{HB}}=$ $1.57\pm 0.03$. Note that allowing the photon indices to be free for each observation, does not improve the spectral fit for none of the two baseline models.}

{We then add the the reflection component ({\tt pexmon} within {\sc xspec}) and refit the data for both the uniform and the partial-covering scenarios. We let $\rm{\Gamma_{pex}}$ and the {\tt pexmon} normalization linked between observations, with $\rm{\Gamma_{pex}}$ fixed to $\rm{\Gamma_{HB}}$. For $\rm{M_1}$ we obtain a $\rm{\chi^2/\rm{dof} = 1060.1/1048}$, {$  \rm{N_H} = 11.2\pm0.4 \times 10^{22} \ cm^{-2}$} and $\rm{\Gamma_{HB}}=$ $1.56\pm 0.04$. For $\rm{M_2}$ we obtain $\rm{\chi^2/\rm{dof} = 1025.4/1047}$, with { $\rm{N_H} = 22.1\pm1.9 \times 10^{22} \ cm^{-2}$} and $\rm{\Gamma_{HB}}=$ $1.69\pm 0.05$. We also checked that letting free $\rm{\Gamma_{HB}}$ or $\rm{\Gamma_{pex}}$ does not produce statistically better results.}  

{This results in a $\Delta \chi^2$ of 94 and 74 for $\rm{M_1}$ and $\rm{M_2}$, respectively, for just one extra dof, which is highly significant (the F-test implies a probability of $\rm{\sim10^{-21}}$ and $\rm{10^{-17}}$, respectively, for this to happen by chance). Therefore, the reflection component is highly significant. In the subsequent analysis we fix the reflection component parameters obtained with \emph{Suzaku} and \emph{NuSTAR} data for the \emph{XMM}-Newton observations, which are less sensitive to the reflection component.}

\subsection{Hard band spectral fits}\label{hb-fit}

{The hard energy band for all the spectra are now simultaneously fitted using the two baseline models described in Eqs. \ref{uniform-absorber} and \ref{partial-covering}, with the addition of the fixed circumnuclear contribution (see Sec. \ref{circumnuclear}). Note that we keep the photon index of N2 equal to that of X5 plus the correction factor 
reported in Sec. \ref{sec:mismatch}.}

{We firstly fit the data to $\rm{M_1}$ and $\rm{M_2}$ with all the parameters linked between observations. We find $\rm{\chi^2/dof = 2910.7/2819}$ and $\rm{\chi^2/dof = 2821.4/2818}$ for $\rm{M_1}$ and $\rm{M_2}$, respectively. We then allow absorption variations in both cases, obtaining $\rm{\chi^2/dof = 2779.6/2813}$ and $\rm{\chi^2/dof = 2757.3/2812}$ for $\rm{M_1}$ and $\rm{M_2}$, respectively. This improvement is highly significant with F-test probability of $\rm{10^{-25}}$ and $\rm{10^{-12}}$, respectively. In the case of the partial-covering model, $\rm{M_2}$ we find $\rm{C_f = 0.917 \pm 0.003}$.} 

{Note that we also try to firstly vary the photon indices rather than the absorption. In this case, we obtain $\rm{\chi^2/dof = 2834.7/2813}$ and $\rm{\chi^2/dof = 2761.9/2812}$ for $\rm{M_1}$ and $\rm{M_2}$, respectively. The $\epsilon_2$ test implies that either $\rm{N_H}$ or $\rm{\Gamma_{HB}}$ variations can explain the observed spectral variability in the case of the $\rm{M_2}$ model equally well. However, in the case of $\rm{M_1}$,  $\epsilon_2 \sim 10^{-12}$. This result implies that it is $\sim 10^{12}$ times more probable for the $\rm{N_H}$ to be variable, than $\rm{\Gamma_{HB}}$, for the given spectra. In any case, we will further investigate the $\rm{N_H}$ and/or $\rm{\Gamma_{HB}}$ variations when we fit the full-band spectra (see below).}

\subsection{Full band spectral fits}\label{sec:fullband}

\begin{figure}
    \includegraphics[width=1\columnwidth]{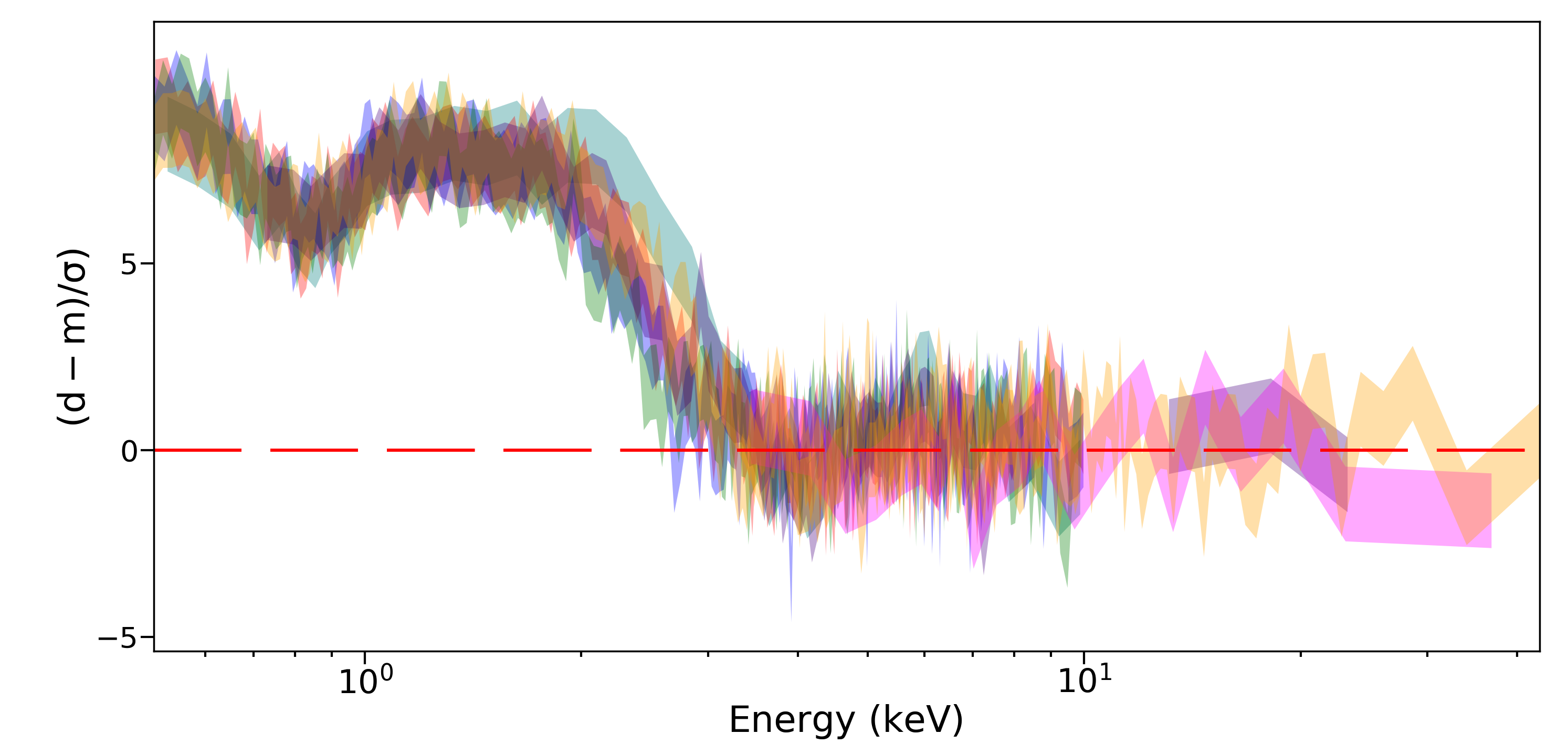}
    \includegraphics[width=1\columnwidth]{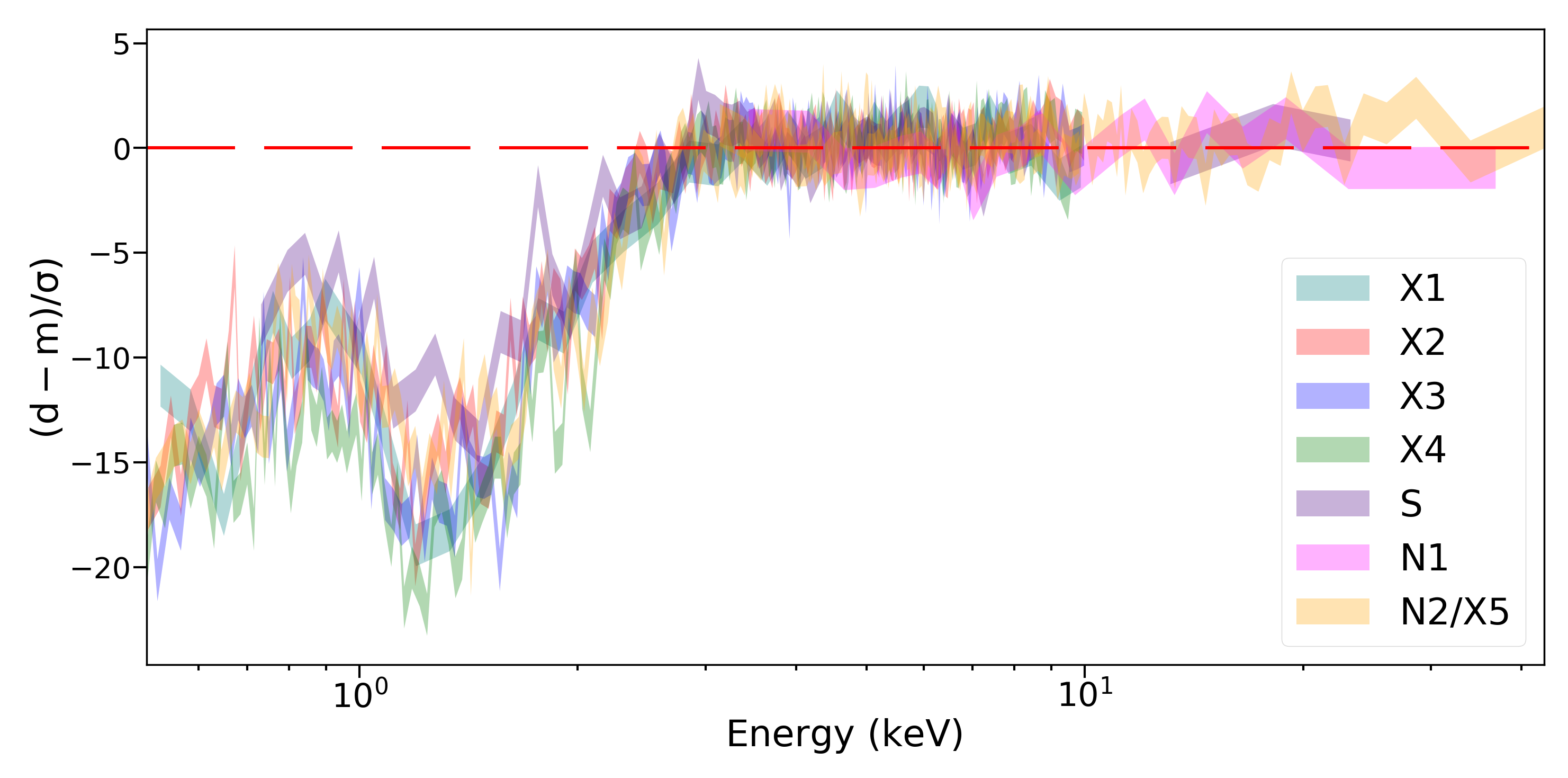}
    \caption{Full band (0.5-50 keV) residuals when extrapolating the spectra below 3\,keV for $\rm{M_1}$ (uniform absorber, top) and $\rm{M_2}$ (partial-covering absorber, bottom).}
    \label{fig:softresiduals}
\end{figure}

{We use the best-fit obtained for the hard-band spectral fit assuming absorption variations to extrapolate it and compare it to the data below 3 keV. Fig.\,\ref{fig:softresiduals} shows the residuals when the hard-band best-fit models to the uniform absorber model, $\rm{M_1}$, and the partial-covering absorber model, $\rm{M_2}$, are extended to low energies. In both cases there is a mismatch between the data and the models. {An excess (deficit) of flux below $\rm{\sim 3}$\,keV is observed for the baseline model $\rm{M_1}$ ($\rm{M_2}$)}. Both are dependent on the observations, although this effect is better seen for the partial-covering absorber model $\rm{M_2}$.}

{As a reminder we keep normalization and photon index of the reflection component to those obtained for the hard-band spectral fits using \emph{NuSTAR} and \emph{Suzaku} data, because this component is better restricted using high energies.}
\begin{table*}
\def\arraystretch{1.5}

\begin{tabular}{cccccccc}

\hline
 OBS & \multicolumn{3}{c}{$\rm{M_1}$} & & \multicolumn{3}{c}{$\rm{M_2}$} \\ \cline{2-4}  \cline{6-8}
 & $\rm{N_{H_{intr}}}$  &  $\rm{\Gamma_{HB}}$ & $\rm{\log F (2-10\,keV)}$ & & $\rm{N_{H_{intr}}}$  &  $\rm{\Gamma_{HB}}$ & $\rm{\log F (2-10\,keV)}$ \\  
  & $10^{22} \ (\rm{cm^{-2}})$ & & $\rm{erg \ s^{-1} \ cm^{-2}}$ & & $10^{22} \ (\rm{cm^{-2}})$ & & $\rm{erg \ s^{-1} \ cm^{-2}}$ \\ \hline \hline

X1 & $10.82_{-2.11}^{+2.31}$ & $1.16 \pm {0.20}$ & $-11.34 \pm {0.03}$ & & $17.38_{-1.30}^{+1.42}$ & $1.62 \pm {0.03}$  & $-11.21 \pm {0.02}$  \\ 

X2 & $7.93_{-0.45}^{+0.47}$ & $1.20 \pm {0.07}$ & $-11.28 \pm {0.01}$ & & $10.75_{-0.34}^{+0.33}$ & $1.60 \pm {0.02}$  & $-11.21 \pm {0.01}$ 
  \\ 
S & $11.24_{-0.76}^{+0.80}$ & $1.46 \pm {0.08}$ & $-11.14 \pm {0.01}$ & & $12.64_{-0.55}^{+0.57}$ & $1.59 \pm {0.03}$  & $-11.09 \pm {0.01}$  \\ 
X3 & $8.57_{-0.40}^{+0.42}$ & $1.41 \pm {0.06}$ & $-11.18 \pm {0.01}$ & & $9.55_{-0.27}^{+0.26}$ & $1.56 \pm {0.02}$  & $-11.15 \pm {0.01}$  \\ 
X4 & $9.59_{-0.38}^{+0.39}$ & $1.52 \pm {0.06}$ & $-11.14 \pm {0.01}$ & & $9.72_{-0.25}^{+0.23}$ & $1.52 \pm {0.02}$  & $-11.12 \pm {0.01}$  \\ 
N1  & $11.67_{-1.49}^{+1.57}$ & $1.53 \pm {0.06}$ & $-11.12 \pm {0.03}$ & & $14.01_{-2.19}^{+2.47}$ & $1.54 \pm {0.07}$  & $-11.11 \pm {0.03}$  \\ 
N2/X5 & $13.34_{-0.38}^{+0.39}$ & $1.58 \pm {0.02}$ & $-11.00 \pm {0.01}$ & & $13.54_{-0.33}^{+0.31}$ & $1.51 \pm {0.01}$  & $-11.02 \pm {0.01}$  \\ 

\hline
\end{tabular}
\caption{\textit{Best-fit parameters for M1 (left) and M2 (right). First (fourth) column is the column density in units of $10^{22} \rm{cm^{-2}}$, second (fifth) column is the photon index and third (sixth) column is the flux of the intrinsic, unabsorbed continuum.} } 
\label{final-table}
\end{table*}
\subsubsection{Uniform absorber $\rm{M_1}$}

{In the case of the uniform absorber baseline model $\rm{M_1}$, we add an extra power-law component, with a photon index named $\rm{\Gamma_{SB}}$. We test two scenarios for $\rm{\Gamma_{SB}}$:}

\begin{enumerate}
  
    \item {$\rm{\Gamma_{SB}}$ is linked to that of the reflection ($\rm{\Gamma_{pex}=1.56}$), which would suggest that the emission associated with this soft band power-law component smears out through a complex region. In this case we obtain $\rm{\chi^2/dof = 4226.6/4054}$, and $\rm{N_{H}=[9.4-14.9]\times 10^{22}cm^{-2}}$ and $\rm{\Gamma_{HB}= 1.53 \pm 0.02}$.}
    
    \item {$\rm{\Gamma_{SB}}$ is free to vary among the observations, being also independent to $\rm{\Gamma_{HB}}$. In this case, $\chi^2/\rm{dof} = 4215.1/4047$ and a range of $\rm{N_{H}=[9.1-15.9]\times 10^{22}cm^{-2}}$ with $\rm{\Gamma_{HB}=1.52\pm 0.02}$, and $\rm{\Gamma_{SB} = [1.4-1.8]}$. Note that in this case, $\rm{\Gamma_{SB}}$ is not constrained for the N1 observation due to the lack of data below 3 keV. This fit is no better than (\textit{i}).}
\end{enumerate}

{We also test if intrinsic continuum photon-index variations between observations are required when the full-band spectra are analysed. So, when $\rm{\Gamma_{SB}}$ is linked to that of the reflection ($\rm{\Gamma_{pex}=1.56}$), and we let  $\rm{\Gamma_{HB}}$ free to vary, the fit improves significantly with $\rm{\chi^2/dof = 4188.8/4048}$. This is statistically better than without intrinsic continuum photon index variations with a F-test probability of $\rm{5\times 10^{-6}}$. In this case, we obtain a range of $\rm{N_{H}=[8.6-13.3]\times 10^{22}cm^{-2}}$ and $\rm{\Gamma_{HB}}= [1.2-1.6]$.} 

With this scenario as the best-fit model for $\rm{M_1}$ we add a warm absorber ({\tt absori} within {\sc xspec}) since it has been previously reported to be present in this source \cite[e.g.][]{Brenneman-09}. This model has as free parameters the temperature, the column density ($\rm{N_{H_{warm}}}$), the photon index, the ionization state ($\rm{\xi}$) and the iron abundance. We linked $\rm{N_{H_{warm}}}$ and $\rm{\xi}$, while the photon index is linked to that of the primary continuum, and the rest of the parameters are kept frozen at their default values. {The improvement of the fit is not statistically significant ($\chi^2/\rm{dof} = 4183.5/4046$), which means that, under this baseline model, NGC\,1052 does not present a warm absorber.} 

\subsubsection{Partial-covering absorber $\rm{M_2}$}
\label{sec:partial-covering}

{In the case of the partial-covering baseline model $\rm{M_2}$, we start by adding a warm absorber, {\tt absori}, due to the deficit seen in Fig.\, \ref{fig:softresiduals}. We keep $\rm{N_{H_{warm}}}$ and $\rm{\xi}$ linked. We also assume the photon index of the warm absorber to be linked to $\rm{\Gamma_{HB}}$ previous section (i.e., the spectral slope of the continuum, which we initially keep linked). We obtain a $\rm{\chi^2/dof = 4241.8/4057}$, with $\rm{N_{H}=[9.5-19.7]\times 10^{22}cm^{-2}}$, $\rm{\Gamma_{HB}}= 1.53 \pm 0.01$. As for the warm absorber, we obtain $\rm{\xi} = 167.5_{-26.2}^{+41.7}$ and {$\rm{N_{H_{warm}} = 0.7\pm 0.1 \times 10^{22}cm^{-2}}$}.\\ 

As in the previous sub-section, we investigate whether variations in $\rm{\Gamma_{HB}}$ (i.e., in the intrinsic spectral slope) are necessary. We refitted the data, allowing $\rm{\Gamma_{HB}}$ variations, obtaining $\rm{\chi^2/dof = 4184.3/4051}$, with $\rm{N_{H}=[9.5-17.4]\times 10^{22}cm^{-2}}$ and $\rm{\Gamma_{HB}= [1.5-1.6]}$. The F-test probability ($\rm{4\times 10^{-10}}$) implies that this model is statistically preferred over the previous one. In other words, just like in $\rm{M_1}$, model $\rm{M_2}$ also suggests that continuum $\rm{\Gamma_{HB}}$ variations are statistically significant. We then refitted the data by allowing $\rm{N_{H_{warm}}}$ variations. However, the improvement to the quality of the fit is not significant in this case ($\rm{\chi^2/dof = 4212.73/4051}$).}

\subsubsection{Baseline model comparison}
\label{sec:comparison}
{For $\rm{M_1}$ the best-fit model suggests variations of the photon index of the intrinsic continuum, $\rm{\Gamma_{HB}}$, and in the uniform absorber $\rm{N_{H_{intrin}}}$, while the reflection component is constant. In the soft band, there are variations in the normalization but the photon index of the soft-band power-law is constant.}
This full band spectral modelling is consistent with results of previous spectral modelling of the source \citep[e.g.][]{Gonzalez-09,Hernandez-13,Hernandez-14}

{For $\rm{M_2}$ we find that the photon index ($\rm{\Gamma_{HB}}$) and the $\rm{N_{H_{intrin}}}$ vary as well. The soft band also shows changes associated with the intrinsic continuum, and in addition, an ionized absorber is needed under this scenario. This modelling is also consistent with previous results \citep[e.g][]{Brenneman-09, Connolly-16}.} Similar spectral models have also been used in the study of the X-ray spectra of other LLAGN as well.

\begin{figure*}
\includegraphics[width=2.\columnwidth]{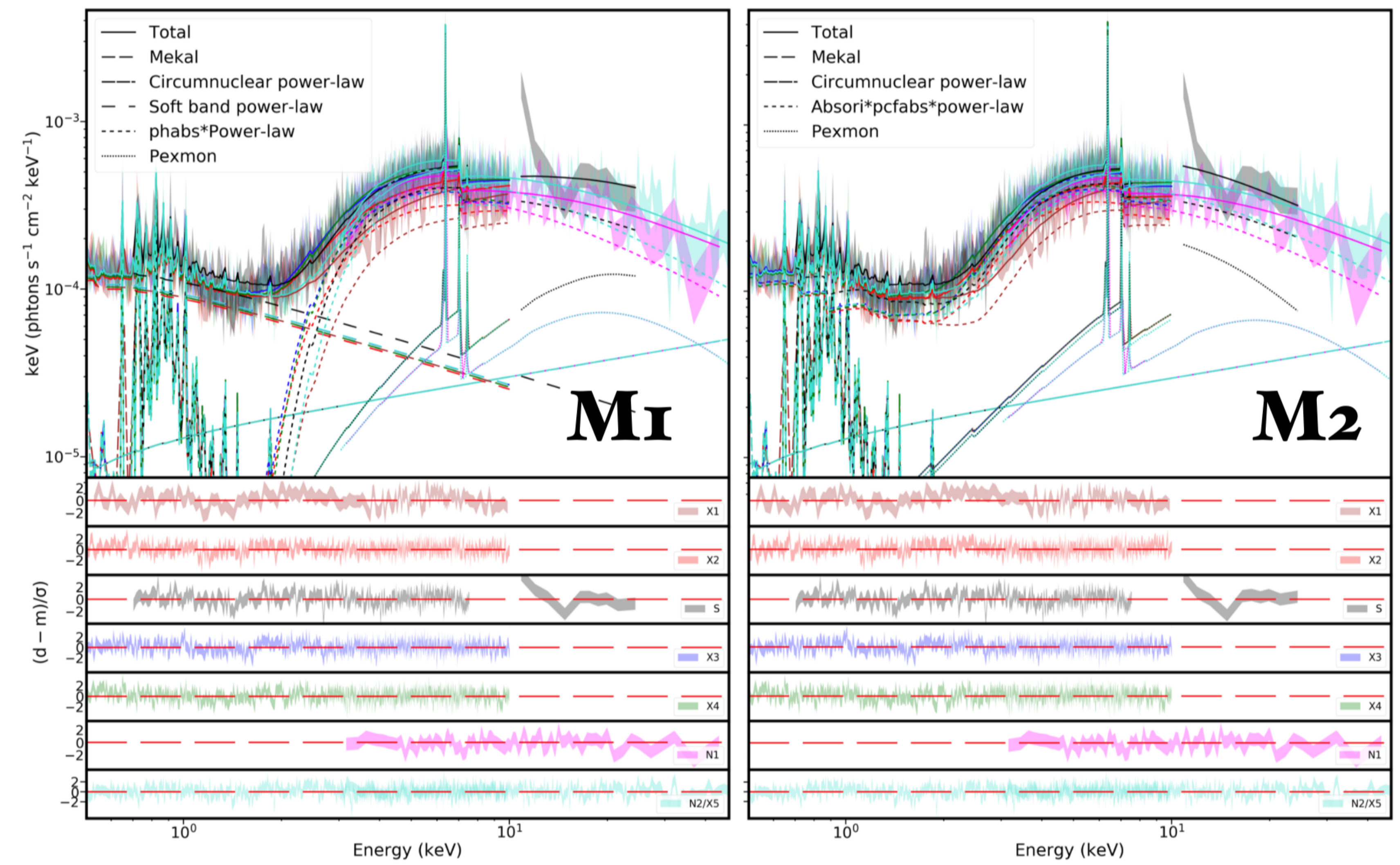}
\caption{Final spectral fits for NGC\,1052 for the uniform absorber model $\rm{M_1}$ (left) and the partial-covering absorber model $\rm{M_2}$ (right). Each observation is shown with different colors: brown, red, blue, and green for the \emph{XMM}-Newton observations X1, X2, X3, and X4, respectively, black for \emph{Suzaku} S, magenta for \emph{NuSTAR} N1, and cyan for the simultaneous \emph{NuSTAR} N2 and \emph{XMM}-Newton X5 observation, respectively. In both panels, the total spectrum, {\tt mekal}, circumnuclear {\tt power-law}, and reflection {\tt pexmon} components are represented as continuous, short-dashed, short-dashed-dotted, and dotted lines. The long-dashed and short-dashed lines in the left panel ($\rm{M_1}$) represent the soft-band {\tt power-law} and uniformly absorbed hard-band {\tt power-law}, respectively. The short-dashed line in the right panel ($\rm{M_2}$) represents the partial-absorbed {\tt power-law}. We also present the residuals as (data-model)/$\sigma$ per observation in the lower panels.}
\label{finalfit}
\end{figure*}

{Both $\rm{M_1}$ and $\rm{M_2}$ fit the data well and they both require changes on the column density and on the photon index. However, when we compare the best spectral fit of the full band baseline model $\rm{M_1}$ with that of baseline model $\rm{M_2}$, by using the $\epsilon_2$ ratio, we find that $\rm{M_2}$ is $\sim 200$ times more likely than $\rm{M_1}$. In other words, statistically speaking, model $\rm{M_2}$ fits the spectra of NGC\,1052 better than model $\rm{M_1}$.} Best-fit parameters for $\rm{M_1}$ and $\rm{M_2}$ are listed in table \ref{final-table} and the best-fit for each model to all data is shown in Fig. \ref{finalfit}. Table \ref{final-table} includes the unabsorbed 2-10 keV fluxes of the intrinsic continuum. These fluxes were estimated using the {\sc xspec} command {\sc cflux}. 

\section{Discussion}\label{discussion}

We have performed a multi-epoch spectral analysis of NGC\,1052 using \emph{XMM}-Newton, \emph{Suzaku}, and \emph{NuSTAR}. {We have tried two baseline models to fit all spectra simultaneously. The $\rm{M_1}$ model assumes that} the nuclear emission is composed by an absorbed continuum plus a reflection component for the hard band, and a power-law component for the soft band: 
\begin{multline}
M_{1} =  {\tt phabs_{\rm{Gal}}}*({\tt zpowerlw} + \\
{\tt phabs_{\rm{intr}}}*{\tt zcutoffpl} + {\tt pexmon} )
\end{multline}

{The $\rm{M_2}$ model assumes a partially absorbed intrinsic continuum which leaks part of the continuum emission towards soft energies. These model also includes an ionized absorber:}
\begin{multline}
M_{2} =  {\tt phabs_{\rm{Gal}}}*\\
({\tt absori* pcfabs_{\rm{intr}}}*{\tt zcutoffpl} + {\tt pexmon} )
\end{multline}

Circumnuclear extended emission seen with \textit{Chandra} data is added to the two models. {Although both models fit the data well (p-value is larger than 0.01 in both cases), model $\rm{M_2}$ is statistically preferred than model $\rm{M_1}$, in the sense that it appears to be 200 more probable.}

The main contributor to the X-rays is the intrinsic power-law continuum. It accounts for $\sim 90\%$ of the intrinsic luminosity in the 2-10 keV band. The X-rays flux is variable with 2-10 keV X-ray flux variations of a factor of $\rm{\sim}$2.0 and $\sim1.5$ between the first and last observation, for $\rm{M_1}$ and $\rm{M_2}$, respectively. In addition to flux variations, data also prefer a scenario where intrinsic absorption and photon index variations are required (see Tab. \ref{final-table} columns 1-2 and 4-5).

For the reflection component (modelled with {\tt pexmon}), the best fit model is one with $\rm{\Gamma_{pex}}$ constant ($\rm{\Gamma_{pex}=1.56 \pm 0.04 / 1.69 \pm 0.05}$ for $\rm{M_1}$ / $\rm{M_2}$). We see no indications of variations in this component. 

As for the soft band, we find for $\rm{M_1}$ model that the soft power-law shows a photon index constant for all observations, equal to $\rm{\Gamma_{pex}}$, and flux variations. For $\rm{M_2}$ model we also find variations associated with the intrinsic continuum which is leaked to lower energies, and absorbed by an ionized steady absorber.

\begin{figure*}
\includegraphics[width=2.0\columnwidth]{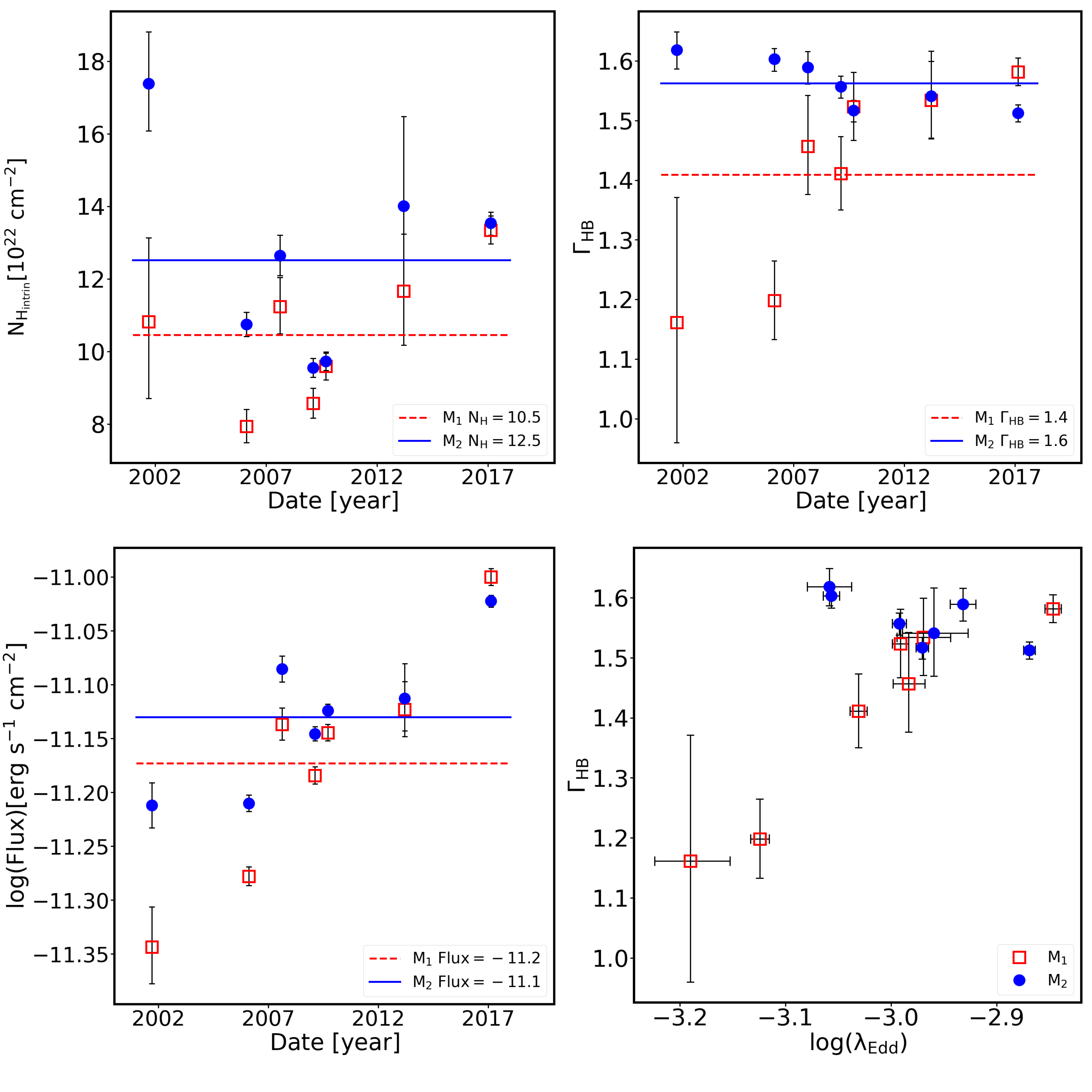}

\caption{In all cases, empty red squares refer to model $\rm{M_1}$ while filled blue circles refer to model $\rm{M_2}$. Top left panel: Best-fit $\rm{N_{H_{intrin}}}$ plotted versus time. Top right panel: Best-fit photon index plotted versus time. Bottom left panel: 2-10 unabsorbed flux versus time. Bottom right panel: Photon index versus Eddington ratio.  In each plot, continuous blue and red dashed lines are the mean values reported in the y-axis for each model. }
\label{contours}
\end{figure*}

\subsection{X-ray reflection} \label{sec:reflection}

We have detected a reflection component in this source which accounts for roughly 10\% of the hard band luminosity. A model including this component is statistically preferred compared to the simpler power-law model using both \textit{NuSTAR} and \textit{Suzaku} data. This is the first time that such component has been detected with high significance in this source. This shows the importance of using broad spectral coverage and multi-epoch observations to disentangle the various components in the X-ray spectra of LLAGN.

It has been previously proposed that this source shows a reflection component, even dominating the emission \citep[e.g.][]{Guainazzi-99} although due to the lack of high energy data it was not constrained until now. \citet{Hernandez-13} modelled the spectra with a thermal component plus two power-laws, not accounting for any reflection component in their model. However, this analysis lacks of data above 10 keV, as the one presented here. \citet{Brenneman-09} established less than 1\% of contribution for the reflection component ($\rm{R < 0.01}$) in the 2007 \emph{Suzaku} observations using {\sc pexrav} while we obtain $\rm{\sim}$10\% using {\sc {\tt pexmon}}. The main difference is that {\sc {\tt pexmon}} self-consistently accounts for both the continuum and the $\rm{Fe \ K_{\alpha}}$ emission line, which is nicely reproduced in our final fit. Thus, we do not see any inconsistency between the amount of flux obtained for the $\rm{FeK_{\alpha}}$ emission line, as claimed by \citet{Brenneman-09}. 

This reflection component is well fitted with a constant photon index and normalization for all the observations. 
This suggests that the reflective material is far enough from the accretion disc with variations dissipated along the path. This strongly support the scenario in which the reprocessor of this emission is the AGN torus, as suggested in other AGN \citep[e.g.][]{Hernandez-15, Balokovic-18}.

\subsection{X-ray intrinsic absorption} \label{sec:absorption}
The left top panel in Fig.\,\ref{contours} shows the best-fit $\rm{N_{H_{intr}}}$ values as a function of time. The red open squares and blue filled circles represent the $\rm{M_1}$ and $\rm{M_2}$ results, respectively. The red dashed and blue continous line represent the mean $\rm{N_{H_{intr}}}$ values in each case. The data clearly show that a single value cannot be representative of all the observations. In both models we note that the combination of \emph{NuSTAR} and \emph{XMM}-Newton data (last square/circle in the plot) gives the most accurate measurement of the $\rm{N_{H_{intr}}}$, reinforcing the need of wide spectral range coverage.

Therefore we are confident on the detection of significant absorption variations in NGC\,1052. Our best-fit models indicate $\rm{N_{H_{intr}}}$ variations along the line of sight with values in the range $\mathrm{[8.6-13.3] \times10^{22} \  cm^{-2}}$ and $\mathrm{[9.5-17.4] \times10^{22} \  cm^{-2}}$ for models $\rm{M_{1}}$ and $\rm{M_{2}}$, respectively. All values are in the regime of Compton-thin sources, consistent with the fact that the flux of reflection component is not dominant in the hard band. 

In the case of the baseline model $\rm{M_2}$ we find a covering factor of $\rm{C_f} \sim 0.9 $, indicating that this is a highly covered source. This value is in agreement with \citet{Connolly-16}, who modelled the spectra of NGC\,1052 with a partial-covering absorber \citep[see also][]{Weaver-99, Hernandez-13}.

The origin of this absorber can be constrained by the fact that it shows absorption variations in scales of months or years. According to model $\rm{M_1}$, the central source is absorbed by a uniform absorber, which can be the case of an out-flowing variable wind. In fact, supporting this scenario, \cite{Cazzoli-18} proposed that this object presents signs of out-flowing winds by using optical 2D spectra. In the case of $\rm{M_2}$, the clumpy absorber can be associated with many clouds, for instance clouds in the BLR. Such clouds do exist in NGC\,1052 \citep{Barth-99}, in contrast to recent studies which claim that such clouds do not exist in LINERs, where it is thought that they disappear due to the low accretion rates \citep[e.g.][]{Elitzur-06, Gonzalez-12}. 
\subsection{X-ray primary continuum, accretion mode and link to the absorption variations} \label{sec:primary}

We found significant spectral slope variations with \ $\rm{\Gamma_{HB}\sim}$1.2 - 1.6 and $\rm{\Gamma_{HB}\sim}$1.5 - 1.6 for $\rm{M_1}$ and $\rm{M_2}$ baseline models, respectively (see Table\,\ref{final-table}). The top right panel in Fig. \ref{contours} shows the time best-fit $\rm{\Gamma_{HB}}$ values as a function of time for both models. It is clear that a single photon index cannot explain all the spectra. These results are consistent with previous works that have studied variability for this source. \cite{Brenneman-09}, for instance, found  $\rm{\Gamma_{HB}} \sim 1.5$ using \emph{Suzaku}, \cite{Hernandez-13} found values from $1.2-1.4$ using \emph{XMM}-Newton and \emph{Chandra}, and \citet{Connolly-16} found values from $1.4- 1.7$ using \textit{Swift} observations.

{We also found intrinsic continuum variations (2-10 keV intrinsic continuum fluxes in Table\,\ref{final-table}). The bottom left panel in Fig.\,\ref{contours} shows the time variability of the 2-10 keV intrinsic flux. It is worth noticing that, irrespective of the model used, it seems that the first two epochs caught the source in a low state, characterized by low intrinsic luminosity.  Using the flux measurements, we can derive the Eddington ratio, assuming the $\rm{L_{bol}/L_{edd}\propto L_{(2-10\,keV)}/L_{edd}}$. With the conversion factor of 50 \citep[][]{Eracleous-10}, our flux measurements imply a range of $\rm{L_{bol}/L_{edd}=[-3.18,-2.84] \ ([-3.06,-2.86])}$ for $\rm{M_1}$ ($\rm{M_2}$) model.}
The bottom right panel in Figure \ref{contours} shows the photon index versus the Eddington ratio for $\rm{M_1}$ (red) and $\rm{M_2}$ (blue) baseline models. This figure suggests a positive correlation between $\rm{\Gamma_{HB}}$ and the 2-10 keV flux (and hence the Eddington rate) in the case of model $\rm{M_1}$, which is mainly driven by the first two \textit{XMM}-Newton observations, with the lowest flux measurements. {However, this is not the case with the model $\rm{M_2}$ results. The small range of variations in $\rm{\Gamma_{HB}}$, if anything, suggests an anti-correlation between $\rm{\Gamma_{HB}}$ and $\rm{\lambda_{Edd}}$.} This is consistent with previous works. In particular, \citet{Connolly-16} used \emph{Swift} monitoring campaign to establish if a relation for NGC\,1052 existed, with observations from 2007 to 2011. They found an anti-correlation between $\Gamma$ and $\rm{L_{bol}/L_{edd}}$, with a much lower Eddington ratio ($\rm{L_{bol}/L_{edd}\sim 10^{-5}}$). However, the low S/N of \emph{Swift} data might have prevented them to model the spectra with a complex baseline model, as the ones reported here and this might have led to an underestimation of the $\rm{L_{bol}/L_{edd}}$.

Different authors \citep[e.g.][]{Constantin-09, Gu-09, Younes-11, Gultekin-12, Jang-14} have suggested that there are two different trends between the photon index and the Eddington ratio. The spectral slope flattens with increasing $\rm{L_{bol}/L_{edd}}$, up to a certain Eddington ratio. At higher accretion rates, the correlation changes and the spectral slope steepens with increasing $\rm{L_{bol}/L_{edd}}$. The spectral slope versus accretion rate correlation is detected in large samples of AGN but also in individual objects. \cite[e.g][]{Sobolewska-09,Emmanoulopoulos-12}. The exact $\rm{L_{bol}/L_{edd}}$ value at which the turn-over happens is still matter of debate, but it is believed that is close to 0.001 \citep{Hernandez-13, Jang-14}. The difference in the $\Gamma$ versus $\rm{L_{bol}/L_{edd}}$ relations has been claimed to be the result of a difference in the accretion mode, with ADAF operating at the low accretion end \citep{Kawamuro-18}, and standard thin accretion disc operating at higher accretion rates. \\
If there is indeed a positive correlation between $\Gamma$ and $\rm{L_{bol}/L_{edd}}$, this would indicate that NGC\,1052 operates on a radiatively efficient mode. {However, the $\rm{M_2}$ results do not support this possibility, which is what would be expected for a LINER}. In this case, the shallowness of the anti-correlation may indicate that the source has reached the turned-out point, around $\rm{\lambda_{Edd}}\sim 0.001$. It is worth noticing again how the use of \emph{NuSTAR} and \emph{XMM}-Newton simultaneous observations (N2, see Fig.\,\ref{contours}) better restricts both $\rm{N_{H_{intr}}}$ and $\rm{\Gamma_{HB}}$ values. Therefore, new simultaneous \emph{NuSTAR} and \emph{XMM}-Newton observations with the source in different flux states are required to fully understand if there is a link between photon index and intrinsic continuum flux beyond statistical errors.

\subsection{Soft X-ray emission} \label{sec:soft}
{According to model $\rm{M_{1}}$, the observed soft band emission is dominated by a separate power-law component. Such a component is often detected in the spectra of LLAGN \citep[e.g.][]{Gonzalez-06,Gonzalez-09,Hernandez-13}. In the case of $\rm{M_{2}}$ this component is not required and instead, an ionized absorber better represents the soft band emission. This ionized absorber has also been claimed by \citet{Brenneman-09}.}

{For $\rm{M_{1}}$, we find that the slope of the soft band continuum emission is constant, which suggests a large scattering region that may smear out changes of the photon index as it would be expected if the soft band power-law accounts for scattered emission from a large ionized region. However, the normalization is variable, which would indicate an origin for this component close to the central source so that it can feel the intrinsic continuum variations in flux, but the $\rm{\Gamma_{HB}}$ variations. It might be the case of a NLR, which has already been proposed by \cite{Bianchi-06} in a sample of Seyfert 2 AGN \citep[see also][]{Gomez-17}.}  
{On the other hand, in the case of the model $\rm{M_{2}}$, the soft band emission is mainly due to a small part of the   
the continuum which leaks through the clumpy neutral absorber. This model requites the presence of a non-variable warm absorber, which is most likely placed within the torus.} 
\section{Summary}
\label{summary}

We have performed a multi-epoch analysis of the LLAGN NGC\,1052 using \textit{XMM}-Newton, \textit{Suzaku}, and \textit{NuSTAR} observations from 2001 to 2017. We used \emph{Chandra} data to separate nuclear from circumnuclear emission. 
The circumnuclear component is composed by a thermal plus a power-law component. The nuclear spectrum emission consists of either a power-law contributing to the soft band, an absorbed power-law, and a reflection component for the hard band (the $\rm{M_1}$ model) or a partial-covering absorber, also including a reflection component and an ionized absorber (the $\rm{M_2}$ model). 
We have found variations both in the intrinsic continuum flux, photon index, and on the obscuration along the line of sight, with a set of $\rm{\Gamma_{HB}}$ values for the continuum that range from $\rm{1.2 - 1.6}$ ($\rm{1.5-1.6}$) and $\rm{N_H}$ in the ranges $[8.6 - 13.3] \times 10^{22} \ \mathrm{cm^{-2}}$ ($[9.5 - 17.4] \times 10^{22} \ \mathrm{cm^{-2}}$) for $\rm{M_1}$ ($\rm{M_2}$). The reflection component is a steady emission both in flux and shape, fully consistent with reflection in a distant structure, perhaps the torus.

Both models fit the data well, but model $\rm{M_2}$ is statistically preferred as providing a better fit to the data and a smaller range of variations, expected for such an object. In addition, it is difficult to explain the soft-band power-law variations in the case of model $\rm{M_1}$, and the $\rm{\Gamma_{HB}}$ versus $\rm{\lambda_{Edd}}$ correlation, whereas the possible anti-correlation in $\rm{M_2}$ is consistent with previous results, as well as with the general relation between these two parameters in LINERs. In addition, the range of variations is much more feasible in the case of model $\rm{M_2}$. For these reasons, we believe that a partial-covering absorber model can explain better the nature of NGC\,1052.
As a final remark we highlight the importance of the simultaneous \emph{XMM}-Newton and \emph{NuSTAR} observations to restrict at the same time photon index and absorption, after taking care of the wavelength-dependent cross-calibration between \emph{XMM}-Newton and \emph{NuSTAR} observations found. In our particular set of data implies $\rm{\Delta \Gamma = \Gamma_{NuSTAR} - \Gamma_{XMM} = 0.17\pm0.04}$.

\section*{Acknowledgments}

We thank the anonymous referee for her/his useful comments which greatly improved this paper. This project is mainly funded by the DGAPA PAPIIT project IA103118.  This research has made use of the NASA/IPAC Extragalactic Database (NED), which is operated by the Jet Propulsion Laboratory, California Institute of Technology, under contract with the National Aeronautics and Space Administration. This research made use of data obtained from the \textit{Suzaku} satellite, a collaborative mission between the space agencies of Japan (JAXA) and the USA (NASA). This research made use of data obtained from the Chandra Data Archive, and software provided by the \textit{Chandra} X-ray Center (CXC) in the application pack- age CIAO. This research made use of data obtained from the \textit{XMM}-Newton Data Archive provided by the \textit{XMM}-Newton Science Archive (XSA). This research has made use of data and/or software provided by the High Energy Astrophysics Science Archive Research Center (HEASARC), which is a service of the Astrophysics Science Division at NASA/GSFC and the High Energy Astrophysics Division of the Smithsonian Astrophysical Observatory. NOC would like to thank CONACyT scholarship No. 897887, LHG acknowledges support from FONDECYT through grant 3170527, JMG acknowledges support from the Spanish Ministerio de Ciencia, Innovación y Universidades through the grant AYA 016-76682-C3-1-P.






\label{lastpage}
\bsp	
\end{document}